\documentstyle[prd,aps,preprint,tighten,epsfig]{revtex}

\begin{document}

\draft

\title{Tetra-maximal Neutrino Mixing and Its Implications on \\
Neutrino Oscillations and Collider Signatures}
\author{{\bf Zhi-zhong Xing}
\thanks{E-mail: xingzz@ihep.ac.cn}}
\address{Institute of High Energy Physics, Chinese Academy of
Sciences, Beijing 100049, China}

\maketitle

\begin{abstract}
We propose a novel neutrino mixing pattern in terms of only two
small integers $1$ and $2$ together with their square roots and the
imaginary number $i$. This ansatz is referred to as the
``tetra-maximal" mixing because it can be expressed as a product of
four rotation matrices, whose mixing angles are all $\pi/4$ in the
complex plane. It predicts $\theta^{}_{12} = \arctan(2-\sqrt{2})
\approx 30.4^\circ$, $\theta^{}_{13} =
\arcsin[(\sqrt{2}-1)/(2\sqrt{2})] \approx 8.4^\circ$,
$\theta^{}_{23} = 45^\circ$ and $\delta =90^\circ$ in the standard
parametrization, and the Jarlskog invariant of leptonic CP violation
is found to be ${\cal J} = 1/32$. These results are compatible with
current data and can soon be tested in a variety of neutrino
oscillation experiments. Implications of the tetra-maximal neutrino
mixing on the decays of doubly-charged Higgs bosons
$H^{\pm\pm}\rightarrow l^\pm_\alpha l^\pm_\beta$ (for $\alpha, \beta
= e, \mu, \tau$) are also discussed in the triplet seesaw mechanism
at the TeV scale, which will be explored at the upcoming LHC.
\end{abstract}


\newpage


\framebox{\bf 1} ~ Recent solar \cite{SNO}, atmospheric \cite{SK},
reactor \cite{KM} and accelerator \cite{K2K} neutrino experiments
have convincingly verified the hypothesis of neutrino oscillation,
a pure quantum phenomenon which can naturally occur if neutrinos
are massive and lepton flavors are mixed. The mixing of lepton
flavors is described by a $3\times 3$ unitary matrix $V$, whose
nine elements are commonly parametrized in terms of three rotation
angles and three CP-violating phases. Defining three unitary
rotation matrices in the complex (1,2), (1,3) and (2,3) planes as
\begin{eqnarray}
O^{}_{12}(\theta^{}_{12}, \delta^{}_{12}) & = & \left(
\matrix{c^{}_{12} & \hat{s}^*_{12} & 0 \cr -\hat{s}^{}_{12} &
c^{}_{12} & 0 \cr 0 & 0 & 1 \cr} \right) \; , \nonumber \\
O^{}_{13}(\theta^{}_{13}, \delta^{}_{13}) & = & \left(
\matrix{c^{}_{13} & 0 & \hat{s}^*_{13} \cr 0 & 1 & 0 \cr
-\hat{s}^{}_{13} & 0 & c^{}_{13} \cr} \right) \; , \nonumber \\
O^{}_{23}(\theta^{}_{23}, \delta^{}_{23}) & = & \left( \matrix{1 & 0
& 0 \cr 0 & c^{}_{23} & \hat{s}^*_{23} \cr 0 & -\hat{s}^{}_{23} &
c^{}_{23} \cr} \right) \; ,
\end{eqnarray}
where $c^{}_{ij} \equiv \cos\theta^{}_{ij}$ and $\hat{s}^{}_{ij}
\equiv e^{i\delta^{}_{ij}} \sin\theta^{}_{ij}$ (for $1\leq i<j
\leq 3$), we can write out the standard parametrization of $V$
advocated by the Particle Data Group \cite{PDG06} and in Ref.
\cite{FX}:
\begin{eqnarray}
V & = & O^{}_{23}(\theta^{}_{23}, 0) \otimes
O^{}_{13}(\theta^{}_{13}, \delta) \otimes O^{}_{12}(\theta^{}_{12},
0) \otimes P^{}_\nu \nonumber \\
& = & \left( \matrix{ c^{}_{12}
c^{}_{13} & s^{}_{12} c^{}_{13} & s^{}_{13} e^{-i\delta} \cr
-s^{}_{12} c^{}_{23} - c^{}_{12} s^{}_{13} s^{}_{23} e^{i\delta} &
c^{}_{12} c^{}_{23} - s^{}_{12} s^{}_{13} s^{}_{23} e^{i\delta} &
c^{}_{13} s^{}_{23} \cr s^{}_{12} s^{}_{23} - c^{}_{12} s^{}_{13}
c^{}_{23} e^{i\delta} & -c^{}_{12} s^{}_{23} - s^{}_{12} s^{}_{13}
c^{}_{23} e^{i\delta} & c^{}_{13} c^{}_{23} \cr} \right) P^{}_\nu \;
,
\end{eqnarray}
in which $P^{}_\nu ={\rm Diag}\{e^{i\rho}, e^{i\sigma}, 1\}$ is a
diagonal phase matrix which contains two non-trivial Majorana phases
of CP violation. A global analysis of current neutrino oscillation
data yields $30^\circ < \theta_{12} < 38^\circ$, $36^\circ <
\theta_{23} < 54^\circ$ and $\theta_{13} < 10^\circ$ at the $99\%$
confidence level \cite{Vissani}, but three phases of $V$ remain
entirely unconstrained. The on-going and forthcoming neutrino
oscillation experiments will measure $\theta^{}_{13}$ and $\delta$.
On the other hand, the neutrinoless double-beta decay experiments
will help to probe or constrain $\rho$ and $\sigma$.

The observed pattern of neutrino flavor mixing is certainly far
beyond the imagination of many people. For instance, the tri-maximal
neutrino mixing proposed by Cabibbo \cite{Cabibbo},
\begin{equation}
V^{}_{\rm C} \; =\; \sqrt{\frac{1}{3}} \left( \matrix{ 1 & 1 & 1
\cr 1 & \omega & \omega^2 \cr 1 & \omega^2 & \omega \cr} \right )
\;
\end{equation}
with $\omega = e^{i2\pi/3}$ being a complex cube-root of unity
(i.e., $\omega^3 =1$), used to be a vivid ansatz in illustration of
both large flavor mixing and maximal CP violation in the lepton
sector; but it has been ruled out by current experimental data on
neutrino oscillations. A simple modification of $V^{}_{\rm C}$,
\begin{eqnarray}
V^{}_{\rm HPS} & = & V^{}_{\rm C} \otimes O^{}_{13}(\pi/4, \pi) \nonumber \\
& = & Q^{}_l \left( \matrix{ \sqrt{\frac{2}{3}} & \sqrt{\frac{1}{3}}
& 0 \cr -\sqrt{\frac{1}{6}} & \sqrt{\frac{1}{3}} &
\sqrt{\frac{1}{2}} \cr \sqrt{\frac{1}{6}} & -\sqrt{\frac{1}{3}} &
\sqrt{\frac{1}{2}} \cr} \right) Q^{}_\nu \;
\end{eqnarray}
with $Q^{}_l = {\rm Diag} \{1, \omega, \omega^2 \}$ and $Q^{}_\nu
= {\rm Diag} \{1, 1, i\}$, which has been proposed by Harrison,
Perkins and Scott \cite{TB} and referred to as the tri-bimaximal
neutrino mixing matrix
\footnote{Note that this pattern is quite similar to the
democratic neutrino mixing pattern \cite{FX96}, although their
consequences on $\theta^{}_{12}$ and $\theta^{}_{23}$ are quite
different.},
turns out to be favored in today's neutrino phenomenology. To
generate the non-vanishing mixing angle $\theta^{}_{13}$ and
non-trivial CP-violating phases, however, slight corrections to
$V^{}_{\rm HPS}$ have to be introduced \cite{XingTB}. So far a lot
of interest has been paid to the tri-bimaximal mixing pattern and
its viable variations, which can be realized in a number of
neutrino mass models incorporated with certain flavor symmetries
and (or) seesaw mechanisms \cite{Review}.

The salient feature of $V^{}_{\rm HPS}$ is that its entries are all
formed from small integers ($0$, $1$, $2$ and $3$) and their square
roots, which are often suggestive of discrete flavor symmetries in
the language of group theories. Then a natural question is whether
one can construct a different but viable neutrino mixing pattern
with fewer small integers. We find that the answer to this
phenomenologically interesting question is affirmative: we may just
use two small integers $1$ and $2$ together with their square roots
and the imaginary number $i$ to build a neutrino mixing matrix which
is compatible with current neutrino oscillation data. This new
pattern, which will be referred to as the ``tetra-maximal" neutrino
mixing, predicts
\begin{eqnarray}
\theta^{}_{12} & = & \arctan
\left[2\left(1-\sqrt{\frac{1}{2}}\right)\right] \; \approx \;
30.4^\circ \; , \nonumber \\
\theta^{}_{13} & = & \arcsin
\left[\frac{1}{2}\left(1-\sqrt{\frac{1}{2}}\right)\right] \;
\approx \; 8.4^\circ \; , \nonumber \\
\theta^{}_{23} & = & 45^\circ \; ,
\end{eqnarray}
and $\delta = 90^\circ$ together with $\rho=\sigma =-90^\circ$.
Since $\theta^{}_{13}$ is large and $\delta$ is maximal, the
Jarlskog invariant of leptonic CP violation \cite{J} turns out to
be ${\cal J} = 1/32$, which can give rise to appreciable effects
of CP or T violation in long-baseline neutrino oscillations. Thus
the tetra-maximal neutrino mixing scenario is easily testable in a
variety of neutrino oscillation experiments in the near future.

\vspace{0.5cm}

\framebox{\bf 2} ~ Now let us describe how to construct the new
neutrino mixing matrix in terms of $1$, $2$ and $i$. We notice
that the tri-maximal mixing pattern $V^{}_{\rm C}$ can be
decomposed as
\begin{equation}
V^{}_{\rm C} \; =\; P^\prime_l \otimes O^{}_{23}(\pi/4, \pi/2)
\otimes O^{}_{13}(\theta^\prime_{13}, 0) \otimes O^{}_{12}(\pi/4,
0) \; ,
\end{equation}
where $P^\prime_l = {\rm Diag}\{1, -i\omega^2, \omega\}$ and
$\theta^\prime_{13} = \arctan(\sqrt{1/2}) \approx 35.3^\circ$.
Therefore, the tri-bimaximal neutrino mixing matrix $V^{}_{\rm
HPS}$ arises from a product of four rotation matrices in the
complex plane: three of them involve the rotation angle $\pi/4$,
and the fourth involves the rotation angle $\theta^\prime_{13}
\neq \pi/4$. The unique value of $\theta^\prime_{13}$ given above
is crucial to assure that Eq. (6) can successfully reproduce the
form of $V^{}_{\rm C}$ in Eq. (3) and then the form of $V^{}_{\rm
HPS}$ in Eq. (4). Indeed, one happens to obtain $\theta^{}_{12} =
\theta^\prime_{13}$ from $V^{}_{\rm HPS}$. This mysterious angle
has a simple geometric explanation \cite{Lee}: it corresponds to
the angle formed by two unequal diagonals from the same vertex of
a cube.

But here we consider the possibility of $\theta^\prime_{13} =
\pi/4$. In this case, we construct a new neutrino mixing pattern
in terms of four rotation matrices whose mixing angles are all
$\pi/4$:
\begin{eqnarray}
V & = & P^{}_l \otimes O^{}_{23}(\pi/4, \pi/2) \otimes
O^{}_{13}(\pi/4, 0) \otimes O^{}_{12}(\pi/4, 0) \otimes
O^{}_{13}(\pi/4, \pi) \nonumber \\
& = & \frac{1}{2} \left(
\matrix{ 1 + \sqrt{\frac{1}{2}} & 1 & 1 - \sqrt{\frac{1}{2}} \cr
-\sqrt{\frac{1}{2}} \left[1 - i \left(1 -
\sqrt{\frac{1}{2}}\right)\right] & 1 - i \sqrt{\frac{1}{2}} &
\sqrt{\frac{1}{2}} \left[1 + i \left(1 +
\sqrt{\frac{1}{2}}\right)\right] \cr - \sqrt{\frac{1}{2}} \left[1 +
i \left(1 - \sqrt{\frac{1}{2}}\right)\right] & 1 + i
\sqrt{\frac{1}{2}} & \sqrt{\frac{1}{2}} \left[1 - i \left(1 +
\sqrt{\frac{1}{2}}\right)\right] \cr} \right) \; ,
\end{eqnarray}
where $P^{}_l = {\rm Diag}\{1, 1, i\}$. It is clear that $V$ only
contains two small integers $1$ and $2$ together with their square
roots and the imaginary number $i$. Because the mixing angle in
each of the four rotation matrices of $V$ is $\pi/4$, this
neutrino mixing matrix can be referred to as the ``tetra-maximal"
neutrino mixing pattern. Some discussions about the
phenomenological consequence of $V$ are in order.
\begin{enumerate}
\item     Comparing between Eqs. (2) and (7), we can easily obtain
the values of three neutrino mixing angles as already listed in
Eq. (5). It is also straightforward to calculate the Jarlskog
invariant of CP violation from Eq. (7):
\begin{equation}
{\cal J} \; = \; {\rm Im}\left(V^{}_{e2}V^{}_{\mu
3}V^*_{e3}V^*_{\mu 2}\right ) \; =\; \frac{1}{32} \; .
\end{equation}
On the other hand, we obtain ${\cal J} = c^{}_{12} s^{}_{12}
c^2_{13} s^{}_{13} c^{}_{23} s^{}_{23} \sin\delta = \sin\delta/32$
from Eq. (2) with the help of Eq. (5). We are therefore left with
$\sin\delta =1$ or equivalently $\delta = \pi/2$. Note that the
maximal value of ${\cal J}$ can only be achieved from the {\it
unrealistic} tri-maximal neutrino mixing pattern $V^{}_{\rm C}$;
i.e., ${\cal J}^{}_{\rm max} = 1/(6\sqrt{3})$. We find that leptonic
CP violation in the tetra-maximal mixing case is about one third of
${\cal J}^{}_{\rm max}$ (namely, ${\cal J}/{\cal J}^{}_{\rm max} =
3\sqrt{3}/16 \approx 32.5\%$).

\item     To figure out two Majorana CP-violating
phases $\rho$ and $\sigma$, we may redefine the phases of three
charged-lepton fields and three neutrino fields such that
$V^{}_{e1}$, $V^{}_{e2}$, $V^{}_{\mu 3}$ and $V^{}_{\tau 3}$ in
Eq. (7) become real and positive while $\delta = \pi/2$ properly
appears in the other five elements of $V$. This exercise will
yield $\rho = \sigma = -\pi/2$. A more straightforward way to
determine $\rho$ and $\sigma$ is to calculate the effective mass
of the neutrinoless double-beta decay by using Eqs. (2) and (5),
\begin{eqnarray}
\langle m\rangle^{}_{\beta\beta} & = & \left| m^{}_1 c^2_{12}
c^2_{13} e^{2i\rho} + m^{}_2 s^2_{12} c^2_{13} e^{2i\sigma} +
m^{}_3 s^2_{13} e^{-2i\delta} \right| \nonumber \\
& = & \frac{1}{4} \left|m^{}_1 \left(1 +
\sqrt{\frac{1}{2}}\right)^2 e^{2i\rho} + m^{}_2 e^{2i\sigma} +
m^{}_3 \left(1 + \sqrt{\frac{1}{2}}\right)^2 e^{-2i\delta}\right|
\; ,
\end{eqnarray}
and compare this result with the one which can be directly
obtained from Eq. (7). We see no interference or cancellation in
the latter procedure, and thus we simply arrive at $\rho = \sigma
= -\delta$ from Eq. (9). Namely, $\rho = \sigma = -\pi/2$.

\item     The off-diagonal asymmetries of $V$, which may serve as a
simple description of the geometric structure of $V$
\cite{Xing02}, are found to be
\begin{eqnarray}
{\cal A}^{}_1 & \equiv & |V^{}_{e 2}|^2 - |V^{}_{\mu 1}|^2 =
|V^{}_{\mu 3}|^2 - |V^{}_{\tau 2}|^2 = |V^{}_{\tau 1}|^2 -
|V^{}_{e 3}|^2 \; = \; - \frac{1}{4} \left(\frac{1}{4} +
\sqrt{\frac{1}{2}}\right) \; , \nonumber \\
{\cal A}^{}_2 & \equiv & |V^{}_{e 2}|^2 - |V^{}_{\mu 3}|^2 =
|V^{}_{\mu 1}|^2 - |V^{}_{\tau 2}|^2 = |V^{}_{\tau 3}|^2 -
|V^{}_{e 1}|^2 \; = \; - \frac{1}{4} \left(\frac{1}{4} -
\sqrt{\frac{1}{2}}\right) \; ,
\end{eqnarray}
which are about $V^{}_{e1}$-$V^{}_{\mu 2}$-$V^{}_{\tau 3}$ and
$V^{}_{e3}$-$V^{}_{\mu 2}$-$V^{}_{\tau 1}$ axes of $V$,
respectively. More explicitly, ${\cal A}^{}_1 \approx -0.24$ and
${\cal A}^{}_2 \approx +0.11$. Hence $V$ looks more symmetric
about its $V^{}_{e3}$-$V^{}_{\mu 2}$-$V^{}_{\tau 1}$ axis. The
fact of ${\cal A}^{}_1 \neq {\cal A}^{}_2 \neq 0$ in the
tetra-maximal mixing case implies that all the six unitarity
triangles of $V$ in the complex plane are different from one
another, although their areas are all equal to ${\cal J}/2 =
1/64$.

\item     The tetra-maximal neutrino mixing pattern shows an
apparent $\mu$-$\tau$ flavor symmetry, $|V^{}_{\mu i}| =
|V^{}_{\tau i}|$ (for $i=1,2,3$), as one can directly see from Eq.
(7). This result has an interesting implication on the flavor
distribution of ultrahigh-energy cosmic neutrinos at neutrino
telescopes. Given the canonical source of cosmic neutrinos, where
the neutrino flavor composition is
\begin{equation}
\phi^{}_e : \phi^{}_\mu : \phi^{}_\tau \; = \; 1: 2: 0 \;
\end{equation}
due to the pion-muon decay chain arising from energetic $pp$ or
$p\gamma$ collisions \cite{Xing06}, the condition of
$\theta^{}_{23} = \pi/4$ and $\delta = \pi/2$ will lead to an
exact neutrino flavor democracy at a terrestrial neutrino
telescope \cite{XZ}:
\begin{equation}
\phi^{\rm T}_e : \phi^{\rm T}_\mu : \phi^{\rm T}_\tau \; =\; 1: 1:
1 \; .
\end{equation}
Note that such a result can also be obtained from the
tri-bimaximal neutrino mixing pattern $V^{}_{\rm HPS}$, which
provides the condition of $\theta^{}_{13} =0$ and $\theta^{}_{23}
= \pi/4$ \cite{Pakvasa}.
\end{enumerate}
In short, the relatively large values of $\theta^{}_{13}$ and
${\cal J}$ predicted by this tetra-maximal neutrino mixing
scenario makes it easily testable in the forthcoming long-baseline
(reactor and accelerator) neutrino oscillation experiments.

\vspace{0.5cm}

\framebox{\bf 3} ~ In the basis where the mass eigenstates of
three charged leptons coincide with their flavor eigenstates, one
may reconstruct the Majorana neutrino mass matrix $M$ by using the
neutrino mixing matrix $V$ and three neutrino masses $m^{}_i$ (for
$i=1,2,3$):
\begin{equation}
M \; =\; V \left(\matrix{ m^{}_1 & 0 & 0 \cr 0 & m^{}_2 & 0 \cr 0
& 0 & m^{}_3 \cr} \right) V^T \; .
\end{equation}
Taking account of the tetra-maximal neutrino mixing pattern given
in Eq. (7), we find that $M^{}_{e\tau} = M^*_{e\mu}$ and
$M^{}_{\tau\tau} = M^*_{\mu\mu}$ hold. Namely,
\begin{equation}
M \; =\; \left(\matrix{ M^{}_{ee} & M^{}_{e\mu} & M^*_{e\mu} \cr
M^{}_{e\mu} & M^{}_{\mu\mu} & M^{}_{\mu\tau} \cr M^*_{e\mu} &
M^{}_{\mu\tau} & M^*_{\mu\mu} \cr} \right) \; .
\end{equation}
Such a specific texture of $M$, which can give rise to the maximal
CP-violating phase in neutrino oscillations (i.e., $\delta = \pm
\pi/2$), is possible to result from a certain flavor symmetry and
its breaking mechanism \cite{Yasue}. Here we focus our interest on
the magnitudes of $|M^{}_{\alpha\beta}|^2$, because they can in
principle be determined from a number of lepton-number-violating
processes in a given model. After a straightforward calculation, we
obtain
\begin{eqnarray}
|M^{}_{ee}|^2 & = & \frac{1}{16} \left[ \frac{17+12\sqrt{2}}{4}
m^2_1 + m^2_2 + \frac{17-12\sqrt{2}}{4} m^2_3 \right . \nonumber \\
&& \left. + \left(3+2\sqrt{2}\right) m^{}_1 m^{}_2 + \frac{1}{2}
m^{}_1 m^{}_3 + \left(3-2\sqrt{2}\right) m^{}_2 m^{}_3 \right] \;
, \nonumber \\
|M^{}_{e\mu}|^2 & = & \frac{1}{16} \left[ \frac{7+4\sqrt{2}}{8}
m^2_1 + \frac{3}{2}m^2_2 + \frac{7-4\sqrt{2}}{8} m^2_3 \right . \nonumber \\
&& \left. - \frac{3+2\sqrt{2}}{2} m^{}_1 m^{}_2 - \frac{1}{4}
m^{}_1 m^{}_3 - \frac{3-2\sqrt{2}}{2} m^{}_2 m^{}_3 \right] \;
, \nonumber \\
|M^{}_{\mu\mu}|^2 & = & \frac{1}{16} \left[
\frac{33-20\sqrt{2}}{16}
m^2_1 + \frac{9}{4}m^2_2 + \frac{33+20\sqrt{2}}{16} m^2_3 \right . \nonumber \\
&& \left. - \frac{9-10\sqrt{2}}{4} m^{}_1 m^{}_2 - \frac{15}{8}
m^{}_1 m^{}_3 - \frac{9+10\sqrt{2}}{4} m^{}_2 m^{}_3 \right] \;
, \nonumber \\
|M^{}_{\mu\tau}|^2 & = & \frac{1}{16} \left[
\frac{33-20\sqrt{2}}{16}
m^2_1 + \frac{9}{4}m^2_2 + \frac{33+20\sqrt{2}}{16} m^2_3 \right . \nonumber \\
&& \left. + \frac{15-6\sqrt{2}}{4} m^{}_1 m^{}_2 + \frac{17}{8}
m^{}_1 m^{}_3 + \frac{15+6\sqrt{2}}{4} m^{}_2 m^{}_3 \right] \; .
\end{eqnarray}
It is then easy to verify
\begin{equation}
\sum_\alpha |M^{}_{\alpha\alpha}|^2 + \sum_{\alpha \neq \beta}
|M^{}_{\alpha\beta}|^2 \; = \; \sum^3_{i=1} m^2_i \; ,
\end{equation}
where $\alpha$ and $\beta$ run over $e$, $\mu$ and $\tau$. Since
the absolute mass scale of $m^{}_i$ is unknown, we consider three
special patterns of the neutrino mass spectrum allowed by current
neutrino oscillation data: (1) normal hierarchy with $m^{}_1
\approx 0$; (2) inverted hierarchy with $m^{}_3 \approx 0$; and
(3) near degeneracy with $m^{}_1 \approx m^{}_2 \approx m^{}_3$.
Taking $\Delta m^2_{21} = 8.0 \times 10^{-5} ~ {\rm eV}^2$ and
$|\Delta m^2_{32}| = 2.5 \times 10^{-3} ~ {\rm eV}^2$
\cite{Vissani} as typical inputs, we are then able to calculate
$|M^{}_{\alpha\beta}|^2$ for three different neutrino mass
hierarchies by using Eq. (15). Our numerical results for
$|M^{}_{\alpha\beta}|^2$ are listed in TABLE I. Note that $\langle
m\rangle^{}_{\beta\beta} \equiv |M^{}_{ee}|$, the effective mass
of the neutrinoless double-beta decay, is found to be $3.3\times
10^{-3}$ eV (normal hierarchy), $4.8\times 10^{-2}$ eV (inverted
hierarchy) or $m^{}_1$ (near degeneracy) in this tetra-maximal
neutrino mixing ansatz.

The origin of $M$ is of course model-dependent. For simplicity, we
assume that $M$ results from the triplet seesaw mechanism
\cite{SS}. By introducing an $SU(2)^{}_{\rm L}$ Higgs triplet
$\Delta$ into the standard model, we can write out the following
renormalizable Yukawa interaction term:
\begin{eqnarray}
-{\cal L}^{}_\Delta \; = \; \frac{1}{2} \overline{l^{}_{\rm L}}
Y^{}_\Delta \Delta i\sigma^{}_2 l^c_{\rm L} + {\rm h.c.} \; ,
\end{eqnarray}
where
\begin{equation}
\Delta \equiv \left(\matrix{H^- & - \sqrt{2} ~ H^0 \cr \sqrt{2} ~
H^{--} & -H^-}\right) \; .
\end{equation}
Note that $\Delta$ can also couple to the standard-model Higgs
doublet $H$ and thus violate lepton number by two units \cite{Zhou}.
When the neutral components of $H$ and $\Delta$ acquire their vacuum
expectation values $\langle H \rangle \equiv v/\sqrt{2}$ and
$\langle \Delta \rangle \equiv v^{}_\Delta$, respectively, the
electroweak gauge symmetry is spontaneously broken and the resultant
Majorana neutrino mass matrix reads $M = Y^{}_\Delta v^{}_\Delta$. A
clear signature of the triplet seesaw mechanism is the existence of
doubly-charged Higgs bosons $H^{\pm\pm}$. If the mass scale of
$\Delta$ is of ${\cal O}(1)$ TeV, then $H^{\pm\pm}$ can be produced
at the LHC via the Drell-Yan process $q\bar{q} \rightarrow \gamma^*,
Z^* \rightarrow H^{++} H^{--}$ or through the charged-current
process $q\bar{q}^\prime \rightarrow W^* \rightarrow
H^{\pm\pm}H^{\mp}$. Note that the masses of $H^{\pm\pm}$ and $H^\pm$
are expected to be nearly degenerate in a class of triplet seesaw
models \cite{SS,Zhou,Triplet}, and thus only $H^{\pm\pm}\rightarrow
l^{\pm}_\alpha l^{\pm}_\beta$ (for $\alpha, \beta = e, \mu, \tau$)
and $H^{\pm\pm} \rightarrow W^\pm W^\pm$ decay modes are
kinematically open. Note also that the leptonic channel
$H^{\pm\pm}\rightarrow l^{\pm}_\alpha l^{\pm}_\beta$ becomes
dominant when $v^{}_\Delta < 1$ MeV is taken \cite{Triplet}.
Therefore, we concentrate on the same-sign dilepton events of
$H^{\pm\pm}$, which signify the lepton number violation and serve
for the cleanest collider signatures of new physics \cite{KS}. The
branching ratio of $H^{--}\rightarrow l^-_\alpha l^-_\beta$ turns
out to be
\begin{equation}
{\cal B}(H^{--}\rightarrow l^-_\alpha l^-_\beta) \; =\; \frac{2}{1 +
\delta^{}_{\alpha \beta}} \cdot
\frac{|M^{}_{\alpha\beta}|^2}{\displaystyle\sum^3_{i=1} m^2_i} \; ,
\end{equation}
which is completely determined by the values of $m^{}_i$ and $V$.
Taking account of Eq. (15), we can estimate the magnitude of
${\cal B}(H^{--}\rightarrow l^-_\alpha l^-_\beta)$ for three
special patterns of the neutrino mass spectrum chosen above. Our
numerical results are listed in TABLE II. The measurement of these
lepton-number-violating decay modes at the LHC will help test the
tetra-maximal neutrino mixing scenario and distinguish it from
other neutrino mixing patterns \cite{LL} in the TeV-scale triplet
seesaw mechanism.

\vspace{0.5cm}

\framebox{\bf 4} ~ Motivated by the principle of simplicity, we
have proposed a novel neutrino mixing pattern in terms of only two
small integers $1$ and $2$ together with their square roots and
the imaginary number $i$. Different from the tri-bimaximal mixing
scenario, our tetra-maximal mixing scenario can accommodate both
non-vanishing $\theta^{}_{13}$ and large CP violation. Its
explicit predictions include $\theta^{}_{12} = \arctan(2-\sqrt{2})
\approx 30.4^\circ$, $\theta^{}_{13} =
\arcsin[(\sqrt{2}-1)/(2\sqrt{2})] \approx 8.4^\circ$,
$\theta^{}_{23} = 45^\circ$, $\delta =90^\circ$, $\rho=\sigma
=-90^\circ$ and ${\cal J} = 1/32$, which are compatible with
current data and can soon be tested in a variety of neutrino
oscillation experiments. We have also illustrated possible
implications of the tetra-maximal neutrino mixing on collider
signatures by taking account of the TeV-scale triplet seesaw
mechanism. In particular, the branching ratios of leptonic decays
of doubly-charged Higgs bosons $H^{\pm\pm}\rightarrow l^\pm_\alpha
l^\pm_\beta$ (for $\alpha, \beta = e, \mu, \tau$) have been
calculated for three special patterns of the neutrino mass matrix.
The results are found to be encouraging and interesting.

The flavor symmetry behind the tetra-maximal neutrino mixing
pattern has to be seen. It is always possible to build a specific
neutrino mass model from which such a flavor mixing pattern can be
derived, although this kind of model building usually relies on
some natural or contrived assumptions. All in all, the
tetra-maximal neutrino mixing can shortly be confronted with a
number of precision neutrino experiments and even the LHC. A test
of its many phenomenological consequences is therefore close at
hand.

\vspace{0.5cm}

{\it Acknowledgments:} This work was supported in part by the
National Natural Science Foundation of China.

\newpage

\begin{table}
\caption{The values of $|M^{}_{\alpha \beta}|^2$ (for $\alpha,
\beta = e, \mu, \tau$) for three special patterns of the neutrino
mass spectrum: (1) normal hierarchy with $m^{}_0 \approx 0$; (2)
inverted hierarchy with $m^{}_3 \approx 0$; and (3) near
degeneracy with $m^{}_1 \approx m^{}_2 \approx m^{}_3$, where
$\Delta m^2_{21} = 8.0 \times 10^{-5} ~ {\rm eV}^2$ and $|\Delta
m^2_{32}| = 2.5 \times 10^{-3} ~ {\rm eV}^2$ have typically been
input. Note that $|M^{}_{e\tau}|^2 = |M^{}_{e\mu}|^2$ and
$|M^{}_{\tau\tau}|^2 = |M^{}_{\mu\mu}|^2$ hold for the
tetra-maximal neutrino mixing matrix under discussion.}
\vspace{0.2cm}
\begin{center}
\begin{tabular}{c|ccc}
& \multicolumn{3}{c}{Neutrino mass hierarchy} \\ & $m^{}_1 \approx
0$ & $m^{}_3 \approx 0$ & $m^{}_1 \approx m^{}_2 \approx m^{}_3$ \\
\hline
$|M^{}_{ee}|^2$ (${\rm eV}^2$) & $1.11 \times 10^{-5}$ & $2.34
\times 10^{-3}$ & $m^2_1$ \\ \hline
$|M^{}_{e\mu}|^2$ (${\rm eV}^2$) & $3.21 \times 10^{-5}$ & $2.56
\times 10^{-5}$ & $0$ \\ \hline
$|M^{}_{\mu\mu}|^2$ (${\rm eV}^2$) & $4.64 \times 10^{-4}$ & $5.94
\times 10^{-4}$ & $0$ \\ \hline
$|M^{}_{\mu\tau}|^2$ (${\rm eV}^2$) & $7.96 \times 10^{-4}$ &
$6.46 \times 10^{-4}$ & $m^2_1$ \\ \hline
$\displaystyle \sum^3_{i=1} m^2_i$ (${\rm eV}^2$) & $2.66 \times
10^{-3}$ & $4.92 \times 10^{-3}$ & $3m^2_1$ \\
\end{tabular}
\end{center}
\end{table}

\begin{table}
\caption{The results of ${\cal B}(H^{--}\rightarrow l^-_\alpha
l^-_\beta)$ (for $\alpha, \beta = e, \mu, \tau$) for three special
patterns of the neutrino mass spectrum: (1) normal hierarchy with
$m^{}_0 \approx 0$; (2) inverted hierarchy with $m^{}_3 \approx
0$; and (3) near degeneracy with $m^{}_1 \approx m^{}_2 \approx
m^{}_3$, where $\Delta m^2_{21} = 8.0 \times 10^{-5} ~ {\rm
eV}^2$, $|\Delta m^2_{32}| = 2.5 \times 10^{-3} ~ {\rm eV}^2$ and
the tetra-maximal mixing parameters have typically been input.}
\vspace{0.2cm}
\begin{center}
\begin{tabular}{c|ccc}
& \multicolumn{3}{c}{Neutrino mass hierarchy} \\ & $m^{}_1 \approx
0$ & $m^{}_3 \approx 0$ & $m^{}_1 \approx m^{}_2 \approx m^{}_3$ \\
\hline
${\cal B}(H^{--}\rightarrow e^-e^-)$ & $0.42\%$ & $47.56\%$ &
$33.33\%$ \\ \hline
${\cal B}(H^{--}\rightarrow e^-\mu^-)$ & $2.41\%$ & $1.04\%$ & $0$
\\ \hline
${\cal B}(H^{--}\rightarrow e^-\tau^-)$ & $2.41\%$ & $1.04\%$ &
$0$ \\ \hline
${\cal B}(H^{--}\rightarrow \mu^-\mu^-)$ & $17.44\%$ & $12.07\%$ &
$0$ \\ \hline
${\cal B}(H^{--}\rightarrow \mu^-\tau^-)$ & $59.85\%$ & $26.26\%$
& $66.67\%$ \\ \hline
${\cal B}(H^{--}\rightarrow \tau^-\tau^-)$ & $17.44\%$ & $12.07\%$
& $0$ \\
\end{tabular}
\end{center}
\end{table}

\end{document}